**Quantum oscillation and nontrivial transport in the Dirac Semimetal $Cd_3As_2$ nanodevice**

Haiyang Pan[1,a)], Kang Zhang[2,a)], Zhongxia Wei[1], Bo Zhao[1], Jue Wang[3], Ming Gao[2], Li Pi[4], Min Han[3], Fengqi Song[1,b)], Xuefeng Wang[2,b)], Baigeng Wang[1], Rong Zhang[2]

[1]National Laboratory of Solid State Microstructures, Collaborative Innovation Center of Advanced Microstructures, and College of Physics, Nanjing University, Nanjing, 210093, P. R. China

[2]National Laboratory of Solid State Microstructures, Collaborative Innovation Center of Advanced Microstructures, and School of Electronic Science and Engineering, Nanjing University, Nanjing, 210093, P. R. China

[3]National Laboratory of Solid State Microstructures, Collaborative Innovation Center of Advanced Microstructures, and Department of Material Science and Engineering, Nanjing University, Nanjing, 210093, P. R. China

[4]Hefei National Laboratory for Physical Sciences at the Microscale, University of Science and Technology of China, Hefei, 230026, P. R. China

---

a)Pan H and Zhang K. contribute equally to this work.

b)Author towhom correspondence should be addressed. Electronic mail: songfengqi@nju.edu.cn; xfwang@nju.edu.cn. Fax:+86-25-83595535

**ABSTRACT:**

Here we demonstrate the Shubnikov de Haas oscillation in high-quality $Cd_3As_2$ nanowires grown by a chemical vapor deposition approach. The dominant transport of topological Dirac fermions is evident by the nontrivial Berry phase in the Landau Fan diagram. The quantum oscillations rise at a small field of 2 Tesla and preserves till up to 100K, revealing a sizeable Landau level gap and a mobility of over 2000 $cm^2/V^{-1}s^{-1}$. The angle-variable oscillations indicates the isotropy of the bulk Dirac transport. The large estimated mean free path appeals the one-dimensional transport of Dirac semimetals.

As the three-dimensional (3D) analogue of graphene,[1] the 3D topological Dirac semimetals (DSM) host the topologically-protected Dirac points, which connect the conduction and valence bands in the Brillouin zone (BZ) and present linear dispersion in all directions around the critical points.[2,3] The quasiparticles may transport in a free-of-backscattering way in the solids of DSMs, leading to various novel physics such ultrahigh room-temperature mobility, fast electronic response and linear unsaturated magnetoresistance.[4-9] Intriguingly, the mean free path can reach hundreds of micrometers at low temperatures.[4] This casts light on a new candidate of the long-desired one dimensional transport, in which the interesting physics of Luttinger liquid and ballistic ejection have been expected. [10,11]

Intense efforts have been made on the confined transport of DSM solids including both the size reduction and mobility improvement.[12-18] However, the sample quality is often suppressed while reducing the lateral sizes, leading to suppressed carrier mobility. For example, $Cd_3As_2$ is one of the promising prototypes of DSM, which was predicted to be DSM as early as 2014 and subsequently confirmed by angle-resolved photoemission spectroscopy.[19-24] A pair of Dirac points are seen in the vicinity of Γ point along $k_z$ with a large Fermi velocity of ~$10^6$m/s.[21] Ultrahigh mobility of $9\times10^6$cmV$^{-1}$s$^{-1}$ and the transport mean free path can reach 200 micrometers at low temperature.[4]. While minimizing the devices by two directions, the mobility is found greatly suppressed. For nanowires with 2 confined dimensions, as the signature of high mobility, Shubnikov de haas (SdH) oscillations have never been detected although ambipolar behavior [12] and negative magnetoresistance in parallel magnetic field are seen.[13-15]. The SdH oscillations can be seen in Cd3As2 nanobelts with only 1 confined dimension, but a trivial Berry phase indicates a non-topological state near the Fermi level.[12] Moreover, the mean free path is also suppressed to 184 nm by 3 orders. Here we demonstrate the SdH oscillation in high-quality $Cd_3As_2$ nanowires grown by a chemical vapor deposition (CVD) approach. The dominant transport of topological Dirac Fermions is evident by the nontrivial Berry phase in the Landau Fan diagram. The estimated transport mean free path appeals the first one-dimensional transport of Dirac semimetals.

The high-quality $Cd_3As_2$ nanowires were grown by a well-controlled CVD method. $Cd_3As_2$ nanowires were grown on Si substrates using $Cd_3As_2$ powder as the precursor in a horizontal tube furnace. The $Cd_3As_2$ powder (J&K Scientific 99.999%) and silicon wafers were placed in a small test tube at first. The substrates were placed close to the open end of the test tube, 15-20cm away from $Cd_3As_2$ powders. Then this test tube was placed inside the quartz tube of the furnace with the $Cd_3As_2$ powder in the center of the furnace and the substrates placed downstream of the argon flow of 100 sccm. Prior to the growth, the tube furnace was pumped to high vacuum and flushed with argon gas several times to remove air and water. Then the temperature of furnace was ramped to 750 °C at a rate 30 °C min$^{-1}$ and kept at 750°C for 30 min with an argon flow as the protective gas. After the furnace was cooled naturally to the room temperature, grey resultant material was found on the surface of the silicon wafers. High quality samples can be selected by electron microscopic inspections.

**Figure 1(a)** shows the scanning electron microscopy (SEM) image of typical $Cd_3As_2$ nanowires which have large aspect ratio and smooth surfaces. The nanowires exhibits the lengths of up to several hundred microns, and the diameter is between 20nm to 300nm. In order to determine the structure of the $Cd_3As_2$ nanowires, we carried out the transmission electron microscopy (TEM) analysis on a typical nanowire. Figure 1(b) shows that the typical nanowire has a diameter of ~50nm. Figure 1(c) is the corresponding high-resolution (HR) TEM image that shows a perfect crystalline structure, in which the 0.73nm interplanar distance indicates the [112] growth direction of the nanowire. The inset displays the Fourier transform of the lattice image which confirms the [112] growth direction. The energy-dispersive X-ray spectroscopy (EDS) acquired from the nanowire in the TEM is shown in Figure 1(d), and the quantitative analysis indicates an atomic ratio of Cd/As = 3:2, which is well consistent with the stoichiometric composition of $Cd_3As_2$. Raman spectroscopy was also carried out on a single $Cd_3As_2$ nanowire, and the result is shown in Figure 1(e). We find two peaks at 189cm$^{-1}$ and 245cm$^{-1}$ respectively, which is consistent perfectly with the α"-Cd3As2 nanocrystal structure reported by S. Wei et al.[25] We transferred the as-grown $Cd_3As_2$ nanowires onto a $SiO_2$/Si substrate to fabricate the field effect transistors(FET). The electrodes patterns are made by E-beam lithographic(EBL) and Ti(5nm)/Au(150nm) was evaporated by E-Beam evaporation(EBE).Then followed by the standard liftoff, the electrodes were fabricated successfully. We fabricated two-electrode devices, where nice contact was achieved and the contact resistance was well below the contribution of the nanowires.[26-28] All the measurements were carried out in the Quantum Design Physical Property Measurement System PPMS-16 systems.

**Figure 2(a)** is the AFM image of our nanowire device, and the diameter is about 180nm. Figure 2(b) shows the temperature dependence of resistance at zero magnetic field. The R-T curve of the nanowire above 50K exhibits a metallic behavior which is consistent with the bulk

R-T, presumably owing to the phonon scattering. Below 50K, the resistance begins to increase as the temperature is decreasing. This is very different from other measurements and this phenomena may be caused by our nanowire's unique quality. Figure 2(c) displays the magnetoresistance under the perpendicular magnetic field ($\theta = 0$) at different temperatures. As we can see, the regular and robust SdH signal is shown without subtracting the background magnetoresistance. This is very different from other SdH observed in $Cd_3As_2$ microstructures.[12,14-16] To gain more insight into the SdH, we subtract a smooth background from the original data at different temperatures. **Figure 3(a)** shows the oscillation amplitudes $\Delta R$ versus 1/B from 2 to 70 K. As depicted in Figure3 (a), the oscillation amplitudes decrease with the increase of the temperature and keeps observable till a high temperature of 70K. A single oscillation frequency F = 28.27 T is obtained from the fast Fourier transform (FFT) Spectra, which corresponds to $\Delta(1/B)= 0.0354$ T$^{-1}$. By using the Onsager relation $F=(\Phi_0/2\pi^2)A_F$, where $\Phi_0$ =h/2e, the cross-sectional area of the Fermi surface normal to the field is $A_F=2.698\times10^{-3}$ Å$^{-2}$. By assuming a circular cross section, a very small Fermi momentum $k_F \approx 0.0293$ Å$^{-1}$ can be extracted, which is very close to the result of 0.0309 Å$^{-1}$ obtained in the microbelts.[16]

To extract more transport parameters of the SdH oscillations, we fit the temperature-dependent SdH oscillation amplitude $\Delta R$ according to the equation [8,12,29] $\Delta R(T)/R(0)=\lambda(T)/\sinh(\lambda(T))$, where the thermal factor is given by $\lambda(T)=2\pi^2 k_B T/(\hbar w_c)$, $k_B$ is the Boltzmann constant, $\hbar$ is the reduced plank constant and $m_{cyc}=E_F/v_F^2$ is the effective cyclotron mass. As shown in Figure 3(b), this equation fits very well and we can obtain the effective cyclotron mass $m_{cyc}=0.052m_e$. By applying the formula $v_F=\hbar k_F/m_{cyc}$, the Fermi velocity

$v_F=6.5\times10^5$ m/s and the Fermi energy $E_F=126$ meV can be acquired. Thereafter, The field dependence of the amplitude of quantum oscillations at fixed temperatures gives access to the Dingle temperature. As $\Delta R(T)/R$ is propositional to $\exp(-2\pi^2 k_B T_D/(\hbar w_c))\lambda(T)/\sinh(\lambda(T))$, where $T_D=\hbar/(2\pi k_B \tau)$ is the Dingle temperature, we can obtain $\tau$ from the slope of the logarithmic plot of $\Delta R(T)\cdot B\cdot\sinh(\lambda(T))/R$ versus $1/B$.[30-32] The relation is shown in Fig3(c), by which $T_D=19.24$K and $\tau=6.32\times10^{-14}$ s are given. The results are close to the value $T_D=23.86$K, $\tau=5.1\times10^{-14}$ in the reported bulk $Cd_3As_2$.[4] Other transport parameters are summarized in Table I.

We then focus on the mean free path of the Dirac fermions. Two types of relaxation periods are recently discussed in the 3D DSM and Weyl semimetals, i.e. the `transport' meanfree time $\tau_{tr}$ extracted by Hall measurement and the quantum mean free time $\tau_Q$ obtained by the SdH oscillations. Due to the coherent protection from the backscattering, $\tau_{tr}$ is at least times of $\tau_Q$. The $\tau_{tr}/\tau_Q$ ratio can even reach $10^4$ in some optimized cases of $Cd_3As_2$.[4] We have got the value of $\tau_Q$ of $6.32\times10^{-14}$ s as stated above. Therefore, the $\tau_{tr}$ of over $10^{-13}$ s is obtained, which leads to the transport mean free path of well over 100nm after considering the Fermi velocity ($l=v_F\tau$). The diameter of the nanowire is 180nm. Our nanowire-based DSM device therefore appeals the 1D transport. The 1D transport might be related to the low-temperature increase of the resistance, which often occurs in the systems with a strong electron-electron interaction.[33] The enhanced electron-electron interaction is some common phenomenon in 1D transport system, as addressed recently in topological systems [34,35]. In addition, no strong localization is seen in our device since $k_f l \gg 1$.

The nontrivial Berry phase demonstrates the topological nature of the device transport.

According to the Lifshitz-Onsager quantization rule,[36] the relation between $1/B$ and $n$ is $1/B = (n + 1/2)(e/hn_s)$. The Landau level (LL) fan diagram is accordingly plotted in Figure 3(d). The maxima of $\Delta R$ are assigned to be the integer indices (solid black) while the minima of $\Delta R$ are plotted by solid blue as the half-integer indices.[29,36] A linear extrapolation of the index plot gives the intercept value -0.59, which is very close to -0.5. This intercept value clear indicates the nontrivial $\pi$ Berry phase, which was observed in the 2D graphene[1,37], topological insulator[36,38], bulkSrMnBi$_2$[39], Rashba semiconductor BiTeI[40], and bulk Cd$_3$As$_2$[8,29]. So we confirmed the quantum transport of topologically-nontrivial Dirac fermions in our Cd$_3$As$_2$ nanowires.

The angular-dependent magnetoresistance measurements are also carried out at 2K. As shown in **Figure4(a)**, the SdH signals are clearly exhibited at every different angles from the perpendicular ($\theta = 0°$) to parallel magnetic field ($\theta = 90°$) after subtracting the background. Figure 4(b) shows the SdH amplitude $\Delta R$ versus $1/B$. We can see that the amplitudes of SdH decrease with the magnetic field change from $\theta = 0°$ to $\theta = 90°$, but still exist in parallel magnetic field. We perform the FFT operation on the SdH oscillations at different tilt angles to obtain the oscillation frequencies F, and then calculate the cross-section areas of the Fermi surface $S_F$ using the equation $F = (\phi_0/2\pi^2)S_F$. The result is displayed in the inset of Figure 4(a) and it clear reveals that the cross-section areas of the Femi surface are almost unchanged when the tilting angles changed from 0° to 90°. This implies a nearly spherical Fermi surface and isotropic $v_F$, which is in good agreement with other transport experimental results.[4,41]

In summary, we synthesized the high-quality Cd$_3$As$_2$ nanowires through the CVD method. The SdH oscillations were observed and the nontrivial $\pi$ Berry's phase revealed the existence of

Dirac fermions in the Cd3As2 naowires for the first time. The long meanfree path appeals1D quantum transport in our nanowires. Our work paves the way to study physical properties and device applications of low dimension DSM materials.

**Acknowledgments**

We gratefully acknowledge the financial support of the National Key Projects for Basic Research of China (Grant Nos:2013CB922103, 2011CB922103), the National Natural Science Foundation of China (Grant Nos: 91421109, 11134005, 11522432, 11274003and 11574288), the Natural Science Foundation of Jiangsu Province (Grant BK20130054), and the Fundamental Research Funds for the Central Universities.

**Figure Captions**

**Figure 1. The synthesized $Cd_3As_2$ nanowires.** (a)The scanning electron microscopic image of the $Cd_3As_2$ nanowires. (b)Transmission electron microscopic (TEM) image of a nanowire with the diameter of 50nm. (c)High resolution TEM image of the nanowire shows a 0.73 nm interplanar spacing of the [112] growth direction. The inset is a Fourier transform of the lattice image. (d) The energy dispersion spectrum shows the atomic ratio of Cd/As = 3:2. (e) Raman spectrum of a $Cd_3As_2$ nanowire with two peaks at 189cm$^{-1}$ and 245cm$^{-1}$ respectively).

**Figure 2. The device and its electrical transport.** (a) The height profile of the white line of the atomic force microscopic image in the inset. The diameter is about 180nm. (b) The temperature dependent resistance. (c) Resistance versus the perpendicular magnetic field B ($\theta = 0$) at different temperatures.

**Figure 3. The Shubnikov de Haas (SdH)oscillations of bulk Dirac Fermions.**(a)The SdH oscillations' amplitude $\Delta R$ as a function of 1/B at different temperatures after subtracting the background. A period is F=28.27 T. (b) The temperature dependence of $\Delta R$ of the second Landau level. The solid line is a fit to the Lifshitz-Kosevich formula and gives the cyclotron effective mass of $0.052m_e$. (c) Dingle plots giving the quantum lifetime$6.32 \times 10^{-14}$ s. (d) LL index plot n versus 1/B. The intercept is-0.59 by taking the maximum and minimum of $\Delta R$ as the integer and half integer, respectively, evidencing the topological transport.

**Figure 4.Angle-dependent SdH analysis** (a) Angle-dependent SdH measurements at 2K. Inset is the cross-section areas of the Fermi surface $S_F$ at different angles. (b) The angle-dependent amplitudes. The curves are vertically displaced for clarity

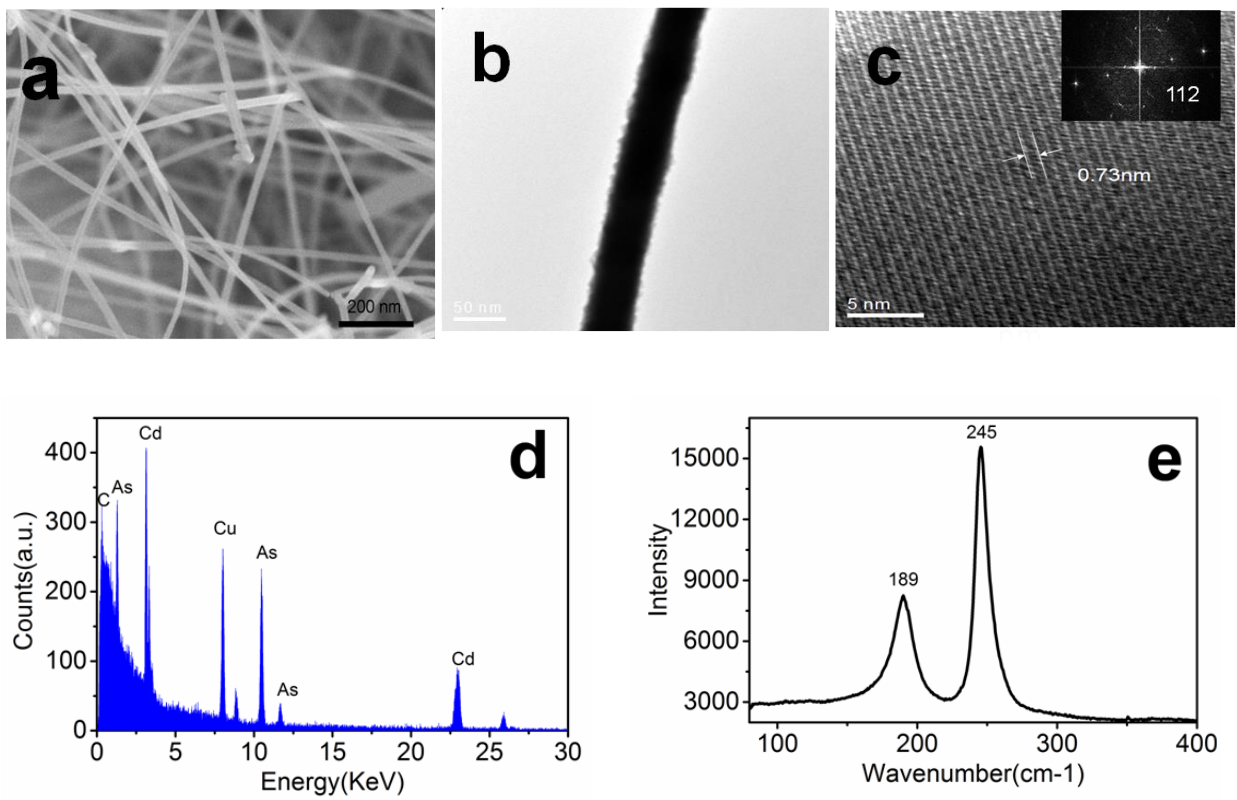

**Figure 1. Pan et al**

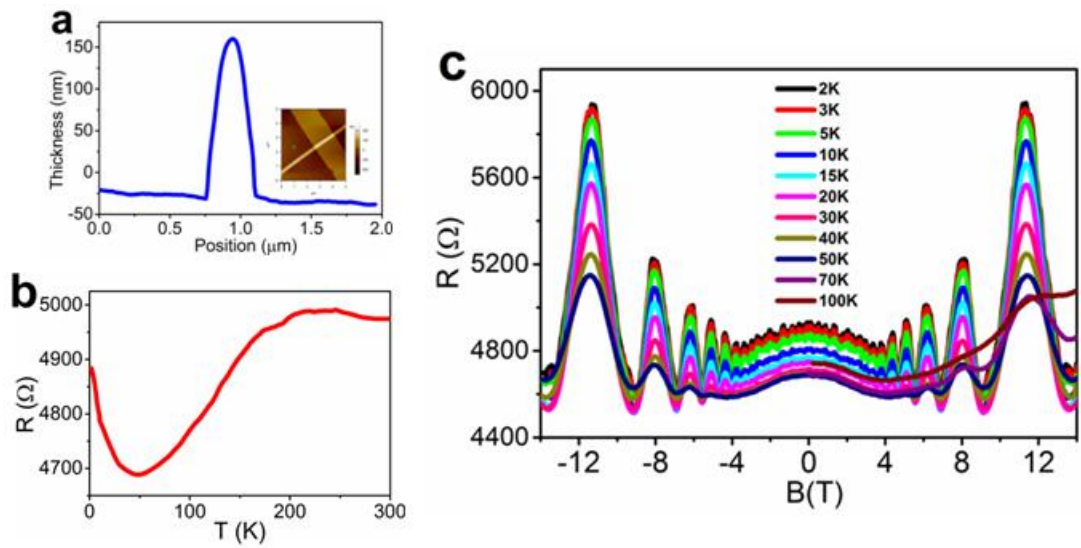

**Figure 2. Pan et al**

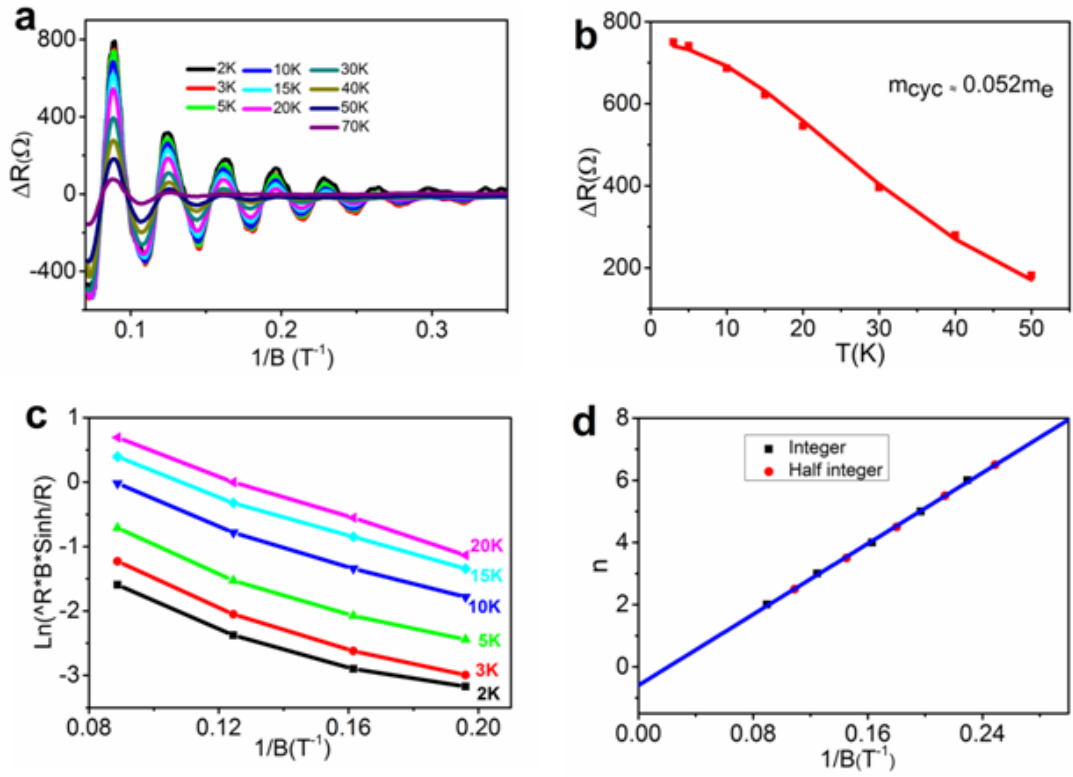

**Figure 3. Pan et al**

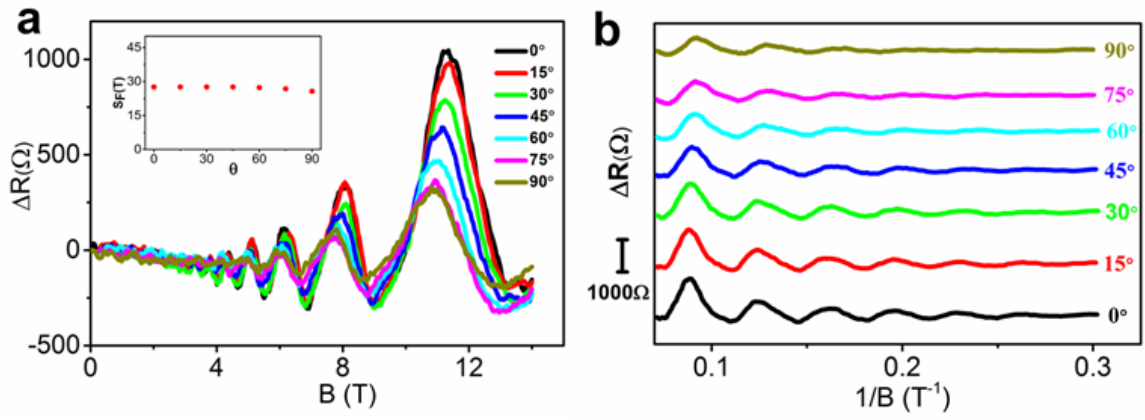

**Figure 4.** Pan et al

| $F_{SdH}$(T) | $S_f(10^{-3}Å^{-2})$ | $k_f(Å^{-1})$ | $m_{cyc}(m_e)$ | $v_F(10^6 m/s)$ | $E_f$(meV) | $t(10^{-13}s)$ | $u_{SdH}(cm^2V^{-1}s^{-1})$ |
|---|---|---|---|---|---|---|---|
| 28.27 | 2.698 | 0.0293 | 0.052 | 0.6526 | 125.9 | 0.632 | 2138 |

TABLE I. Estimated Parameters from the SdH oscillation